\documentclass[prb,showpacs,twocolumn]{revtex4}
\usepackage{graphicx}
\usepackage{dcolumn}
\usepackage{bm}
\usepackage{amsmath}

\makeatletter
\def\ExtendSymbol#1#2#3#4#5{\ext@arrow
6095{\arrowfill@#1#2#3}{#4}{#5}}

\makeatother
\begin{document}

\title{Electronic structure and transport for a laser-field-irradiated
quantum wire with Rashba spin-orbit coupling}

\author{Guanghui Zhou$^{1,2,3}$}
\email{ghzhou@hunnu.edu.cn}
\author{Wenhu Liao$^2$}

\affiliation{$^1$CCAST (World Laboratory), PO Box 8730, Beijing
             100080, China}

\affiliation{$^2$Department of Physics, Hunan Normal University,
             Changsha 410081, China\footnote{Mailing address}}

\affiliation{$^3$International Center for Materials Physics,
             Chinese Academy of Sciences, Shenyang 110015, China}

\begin{abstract}
We investigate theoretically the electronic structure and
transport for a two-level quantum wire with Rashba spin-orbit
coupling (SOC) under the irradiation of an external laser field at
low temperatures. The photon-induced transitions between
SOC-splitted subbands with the same lateral confinement quantum
numbers and between subbands with different confinement quantum
number are expected. Using the method of equation of motion (EOM)
for Keldysh nonequilibrium Green's functions (NGF), we examine the
time-averaged density of states (DOS) and the spin polarized
conductance for the system with photon polarization perpendicular
to the wire direction. Through the analytical analysis and some
numerical examples, the interplay effects of the external laser
field and the Rashba SOC on both the DOS and the conductance of
the system are demonstrated and discussed. It is found that the
external laser field can adjust the spin polarization rate and the
transport of the quantum wire system with some proper Rashba SOC
strengths.
\end{abstract}

\pacs{73.23.-b, 71.70.Ej, 72.25.-b, 78.67.Lt}
\vspace{0.2cm}

\maketitle

\section{Introduction}
In recent years, the effects of SOC in semiconductor mesoscopic
systems have attracted more and more attention since it plays an
important role in the emerging field of spintronics (see recent
review article$^1$ and references therein) since the proposal of
constructing an electronic analog of optic modulator using
ferromagnetic contacts as the spin injector and the detector.$^2$
Many fundamental and interesting phenomena, such as spin
precession,$^{3,4}$ spin accumulation,$^{5,6}$ spin (polarized)
transport$^{7,8}$ and spin Hall effect$^{9,10}$ in the systems
with SOC have been investigated and are under further study now.
Though the SOC has its origin in relativistic effects, it is
regarded vitally in some low-dimensional mesoscopic semiconductor
systems.$^{11,12}$

Usually, two types of SOC are taken into account in the
investigation for systems based on a two-dimensional electron gas
(2DEG) confined in a semiconductor heterostructure. They are
Rashba$^{11}$ and Dresselhaus$^{12}$ SOC, which can be described
by the Hamiltonians
\begin{equation}\label{myeq1}
H_R=\frac{\hbar k_R}{m^*}(\sigma_xp_y-\sigma_yp_x)
\end{equation}
and
\begin{equation}\label{myeq2}
H_D=\frac{\hbar k_D}{m^*}(\sigma_yp_y-\sigma_xp_x),
\end{equation}
respectively, where $m^*$ is the effective electron mass and
${\bf\sigma}=(\sigma_x,\sigma_y,\sigma_z)$ is the vector of Pauli
matrix. The strengths of the two types of SOC are measured in
terms of characteristic wavevectors $k_R$ and $k_D$, respectively.
For some semiconductor based systems (e.g., InAs quantum well),
the Rashba term arising from the structure inversion asymmetry in
heterostructures$^{13,14}$ is roughly one order magnitude larger
than Dresselhaus term which is due to the bulk inversion
asymmetry.$^{15}$ Moreover, the strength of Rashba SOC can be
tuned by external gate voltage,$^{16}$ and its effect on the
systems has been paid more attention, particularly in
quasi-one-dimensional quantum wire system.

Mesoscopic systems with or without external magnetic field in the
presence of SOC have been studied extensively.$^{3-10,17}$ Two
years ago, two independent experiments on the (001)-grown n-type
GaAs multiple quantum well structures had been done by using a
circularly polarized infrared radiation$^{18}$ and the
orthogonally polarized two optical harmonic pulses,$^{19}$
respectively. The spin photoncurrent$^{18}$ and the pure spin
current$^{19}$ due to resonant intersubband transitions have been
observed in the absence of any external magnetic field. Hereafter,
for a single quantum well (2DEG) with SOC irradiated under an
in-plane linearly polarized infrared irradiation, the
spin-dependent density of state (DOS) and the density of spin
polarization has been calculated, and a pure spin current has been
theoretically verified for the system.$^{20}$ Further, a mechanism
for spin-polarized photocurrent generation in a multimode quantum
wire, which is due to the combined effect of the Rashba SOC and a
linearly polarized in-plane microwave irradiation, has been
proposed in the presence of a static in-plane magnetic
field.$^{21}$ On the other hand, the electron transport for a
quantum wire under a time-varying electromagnetic (EM) field
irradiation in the absence of SOC has been analyzed previously by
means of the NGF$^{22}$ and the scattering matrix approach,$^{23}$
respectively. However, a further confined low-dimensional systems,
such as a two-level quasi-one-dimensional quantum wire or
quasi-zero-dimensional quantum dot with SOC under the irradiation
of time-dependent field have been studied rarely.$^{21}$

Mesoscopic two-level system (such as a two-level quantum wire or
quantum dot) is of physically important since it has been proved
to be very useful in describing many aspects of interaction
between EM field and the electrons confined in a heterostructure,
and in application of solid-state electronic device. Therefore, it
is meaningful to investigate the interplay effect between the SOC
and the applied laser filed for a two-level mesoscopic system.

In order to investigate the electronic structure and transport of
a two-level quantum wire with SOC under an intense laser field
irradiation, in this paper we theoretically calculate the
time-averaged DOS and the conductance at the low temperatures for
the system. The interplay effects of different laser frequency and
Rashba SOC strength on the electronic structure and transport are
investigated by using the nonequilibrium Keldysh formulism (NKF).
Through the analysis with a few numerical examples, we find some
characteristics different from those for the similar systems in
previous works.$^{20-23}$

The remainder part of the paper is organized as follows. In Sec.
II, we introduce the model Hamiltonian for our system and give the
NKF straightforwardly, where the time-averaged DOS and the
conductance are calculated analytically. The numerical results and
the discussions are shown in Sec. III. Finally, Sec. IV concludes
the paper.

\section{Model and Formalism}
The NGF approach has been employed in last decades to study a
variety of problems beyond the linear response regime.$^{22}$ Meir
et al$^{24}$ derived a formula for the current through a region of
interacting electrons using the NKF. Changing the one-direction
time axis into a loop with two branches, four Green's functions
depending on the relative positions of $t_a$ and $t_b$ in the loop
can be defined. They are time-ordered, anti-time-ordered and two
distribution Green's functions, respectively. However, only two of
them are independent. We will use the approach of standard
nonequilibrium Keldysh EOM in the present work.

Consider a quasi-one-dimensional system of electrons (a quantum
wire) in the presence of SOC and an external time-dependent laser
field, the model Hamiltonian reads
\begin{equation}\label{myeq3}
H=\frac{{\bf p}^2}{2m^*}+V({\bf r})+H_{so}+V(t),
\end{equation}
where ${\bf r}=(x,y)$ and ${\bf p}=(p_x,p_y)$ are two-dimensional
position and momentum vectors, respectively. The SOC Hamiltonian
$H_{so}$ is generally consisted of $H_R$ and $H_D$, while $V(t)$ is
the potential from the interaction of the external time-dependent
laser field with electrons in the system. The electrons are confined
in the $y$ direction by an infinite square-well potential of width
$a$, i.e.,
\begin{eqnarray}\label{myeq4}
V({\bf r})=\left\{
\begin{array}{l l}
0 & (|y|<a/2)\\
\infty & {(|y|>a/2)},
\end{array}
\right.
\end{eqnarray}
which can eliminate the possibility of SOC due to the effective
electric field coming from the nonuniformity of the confining
potential.$^{25}$

To investigate the effects of SOC and the external field on the
electron transport properties by means of NKF, we rewrite
Hamiltonian (3) in the second-quantized form. For this purpose, we
define that ${a^+_{ks\alpha}}$($a_{ks\alpha}$) creates
(annihilates) an electron with wavevector $k$ and a spin branch
$s$ [$s=\uparrow$ and $\downarrow$, or $+$ and $-$, which is the
spin branch index corresponding to spin-up and spin-down,
respectively. See Eq.(11) for detailed explanation] in mode
$\alpha$ in either the left (L) or the right (R) lead, and
$c_{k_xns}^+$($c_{k_xns}$) creates (annihilates) an electron in
the $n$th transverse mode $|k_x,n,s\rangle$ with wavevector $k_x$
and a spin branch index $s$ in the absence of SOC in the quantum
wire modeled as a two-level ($n=1,2$) system. For convenience, we
choose$^{25}$ the spin polarization axis
$\hat{\bf{n}}=(cos\varphi,sin\varphi)$ to be along the effective
magnetic field due to the SOC for wave propagating in the
$x$-direction such that
\begin{eqnarray}\label{myeq5}
{\begin{array}{*{20}c}|s\rangle\\
\end{array}}=\frac{1}{\sqrt{2}}\left({\begin{array}{*{20}c}
se^{-i\varphi/2}\\
e^{i\varphi/2}
\end{array}}\right)
\end{eqnarray}
with $\varphi\equiv arg[k_D+ik_R]$. With these definitive
operators and spin states, the Hamiltonian for a
laser-field-irradiated two-level quantum wire (connected to two
electrode leads) in the presence of SOC reads
\begin{eqnarray}\label{myeq6}
H&=&\sum_{k,s,\alpha\in{L/R}}\varepsilon_{ks\alpha}a_{ks\alpha}^
+a_{ks\alpha}+\sum_{k_x,n,s}\varepsilon_{ns}(k_x)c_{k_xns}^+c_{k_xns}
\nonumber\\&&+\sum_{k,k_x,n,s,\alpha\in{L/R}}(T_{kk_xns}^{\alpha}
a_{ks\alpha}^+c_{k_xns}+h.c.)\nonumber\\&&+\sum_{k_x,n,n',s,s'}
[\gamma_{nn'}\beta_{ss'}+V_{nsn's'}\cos(\Omega
t)]c_{k_xns}^+c_{k_xn's'},\
\end{eqnarray}
where $\varepsilon_{ks\alpha}$ is the energy level with spin $s$ and
wavevector $k$ in lead $\alpha$, and
\begin{equation}\label{myeq7}
\varepsilon_{ns}(k_x)=\frac{\hbar^2}{2m^*}[(k_x-s k_{so})^2
+(\frac{n\pi}{a})^2]-\Delta_{so}
\end{equation}
is the $n$th sublevel in the wire with $k_{so}=
\sqrt{k^2_R+k^2_D}$ and $\Delta_{so}=\hbar^2 k^2_{so}/{2m}$. In
Hamiltonian (6), the coupling between the electrode leads and the
wire with strength $T_{kk_xns}^{\alpha}$ is represented by the
third term, and the last term describes the adiabatical
electron-photon interaction in the wire$^{22,26}$ and the mixture
of transverse modes due to SOC, where $V_{nsn's'}$ are the dipole
electron-photon interaction matrix elements (MEs) and $\Omega$ the
incident laser frequency. Since the frequencies of interest are in
the range corresponding to wavelengths of the order of hundreds of
nanometers, the spatial variation of the field potential can be
neglected. The SOC mixes the transverse modes through the matrix
element $\gamma_{nn'}\beta_{ss'}$, where
\begin{eqnarray}\label{myeq8}
\gamma_{nn'}=\frac{4nn'}{a(n^2-n'^2)} \left\{
\begin{array}{l l}
(-1)^{\frac{n+n'-1}{2}}&(n\neq n')\\
0 & {(n=n')}
\end{array},
\right.
\end{eqnarray}
and according to the lateral confinement potential$^{25}$
$\beta_{ss'}$ is the element of matrix
\begin{eqnarray}\label{myeq9}
{\begin{array}{*{20}c}\beta=\\
\end{array}} \frac{\hbar^2}{m^*k_{so}}\left[ {\begin{array}{*{20}c}
2ik_Rk_D & k^2_D-k^2_R \\
k^2_R-k^2_D & -2ik_Rk_D
\end{array}} \right].
\end{eqnarray}
In the above Hamiltonian we have neglected electron-electron
interactions since its effect on SOC can be plausibly taken into a
renormalized SOC constant.$^{27}$

For simplicity, we focus on the Rashba SOC effect, i.e., let
$k_D=0$. Furthermore, according to Dyson equation, the coupling
between the electrode leads and the wire only adds a self-energy
term in the NGF, so we firstly calculate the Green's function (GF)
of the quantum wire without considering the electrode leads. In
this case the Hamiltonian of the quantum wire part in the absence
of EM field reads
\begin{eqnarray}\label{myeq10}
H_{wire}&=&\sum_{k_x}[\varepsilon_{1\uparrow}(k_x)c^+_{k_x1\uparrow}c_{k_x1\uparrow}
+\varepsilon_{1\downarrow}(k_x)c^+_{k_x1\downarrow}c_{k_x1\downarrow}\nonumber\\
&+&\varepsilon_{2\uparrow}(k_x)c^+_{k_x2\uparrow}c_{k_x2\uparrow}
+\varepsilon_{2\downarrow}(k_x)c^+_{k_x2\downarrow}c_{k_x2\downarrow}\nonumber\\
&+&\varepsilon_R(c^+_{k_x2\uparrow}c_{k_x1\downarrow}
+c^+_{k_x1\downarrow}c_{k_x2\uparrow}\nonumber\\
&-&c^+_{k_x1\uparrow}c_{k_x2\downarrow}-c^+_{k_x2\downarrow}c_{k_x1\uparrow})],
\end{eqnarray}
where $\varepsilon_R=8\hbar^2k_R/(3m^*a)$. According to Eq.(5),
here the spin-up state $|\uparrow\rangle$ and the spin-down state
$|\downarrow\rangle$ are the linear combination of the eigenstates
of $\sigma_z$
\begin{eqnarray}\label{myeq11}
{\begin{array}{*{20}c}|\uparrow\rangle\\
\end{array}}=\frac{1-i}2\left({\begin{array}{*{20}c}
1\\0
\end{array}}\right)+\frac{1+i}2\left({\begin{array}{*{20}c}
0\\1
\end{array}}\right),\nonumber\\
{\begin{array}{*{20}c}|\downarrow\rangle\\
\end{array}}=-\frac{1-i}2\left({\begin{array}{*{20}c}
1\\0
\end{array}}\right)+\frac{1+i}2\left({\begin{array}{*{20}c}
0\\1
\end{array}}\right),
\end{eqnarray}
with equal probability occupying the real spin-up and spin-down
states in the original spin space, respectively.

For definiteness, we consider the case of the applied incident
laser is polarized along $y$ direction (perpendicular to the wire
direction), hence the diagonal electron-photon interaction MEs are
simply zero in the dipole approximation. Also for simplicity in
calculation we assume phenomenologically that the off-diagonal
electron-photon interaction MEs $V_{1s2s'}=V_{2s1s'}=1.0$ as the
free input parameters (dependent of incident laser intensity) ,
and thus the Hamiltonian (10) becomes
\begin{eqnarray}\label{myeq12}
H'_{wire}&=&\sum_{k_x}\{\varepsilon_{1\uparrow}(k_x)c^+_{k_x1\uparrow}c_{k_x1\uparrow}
+\varepsilon_{1\downarrow}(k_x)c^+_{k_x1\downarrow}c_{k_x1\downarrow}\nonumber\\
&+&\varepsilon_{2\uparrow}(k_x)c^+_{k_x2\uparrow}c_{k_x2\uparrow}
+\varepsilon_{2\downarrow}(k_x)c^+_{k_x2\downarrow}c_{k_x2\downarrow}\nonumber\\
&+&[\frac{1}{2}(e^{i\Omega t}+e^{-i\Omega t})+\varepsilon_R]
(c^+_{k_x1\downarrow}c_{k_x2\uparrow}+c^+_{k_x2\uparrow}c_{k_x1\downarrow})\nonumber\\
&+&[\frac{1}{2}(e^{i\Omega t}+e^{-i\Omega t})-\varepsilon_R]
(c^+_{k_x1\uparrow}c_{k_x2\downarrow}+c^+_{k_x2\downarrow}c_{k_x1\uparrow})\nonumber\\
&+&\frac{1}{2}(e^{i\Omega t}+e^{-i\Omega t})(c^+_{k_x1\uparrow}c_{k_x2\uparrow}
+c^+_{k_x2\uparrow}c_{k_x1\uparrow}\nonumber\\
&+&c^+_{k_x1\downarrow}c_{k_x2\downarrow}+c^+_{k_x2\downarrow}c_{k_x1\downarrow})\}.
\end{eqnarray}
It is seen from Eqs.(10) and (12) that the pure Rashba SOC
induces spin-flip transitions with equal probabilities
(spin-conserving) according to Eq.(6) while the applied laser
field may arouse unequal probability transitions for spin-flip and
spin-conserving due to the interplay between the Rashba SOC and
the field. Our interest is to numerically find which kind of
transitions is favorable for this system.

Next we employe the usually defined retarded GF$^{22,24}$
\begin{eqnarray}\label{myeq13}
G_{nsn's'}^r(t_2,t_1)=\ll c_{k_xns}(t_2),
c_{k_xn's'}(t_1)\gg^r\nonumber\\
=-i\theta(t_2-t_1)\langle\{c_{k_xns}(t_2),
c_{k_xn's'}(t_1)\}\rangle,
\end{eqnarray}
then its corresponding Keldysh EOM is
\begin{eqnarray}\label{myeq14}
i\frac{\partial}{\partial t_2}\ll c_{k_xns}(t_2),c_{k_xn's'}(t_1)\gg^r=\nonumber\\
\delta(t_2-t_1)\langle\{c_{k_xns}(t_2),c_{k_xn's'}(t_1)\}
\rangle\nonumber\\
+\ll[c_{k_xns}(t_2),H],c_{k_xn's'}(t_1)\gg^r.
\end{eqnarray}
Inserting system Hamiltonian (12) into (14) and transforming the
variables to $t_2-t_1$ and $t_1$, and then performing the Fourier
transform to change the variable $t_2-t_1$ into $\omega$, we
finally obtain the diagonal MEs of the two retarded GFs without
the coupling between the electrode leads and the wire
\begin{eqnarray}\label{myeq15}
\{[\omega-\varepsilon_{1/2\uparrow}(k_x)][\omega-\varepsilon_{2/1\downarrow}(k_x)]
-\varepsilon^2_R\}\nonumber\\
\cdot\ll c_{k_x1/2\uparrow},c^+_{k_x1/2\uparrow}\gg_{\omega}^r
=\omega-\varepsilon_{2/1\downarrow}(k_x),
\end{eqnarray}
\begin{eqnarray}\label{myeq16}
\{[\omega-\varepsilon_{1/2\downarrow}(k_x)][\omega-\varepsilon_{2/1\uparrow}(k_x)]
-\varepsilon^2_R\}\nonumber\\
\cdot\ll c_{k_x1/2\downarrow},c^+_{k_x1/2\downarrow}
\gg_{\omega}^r =\omega-\varepsilon_{2/1\uparrow}(k_x),
\end{eqnarray}
\begin{eqnarray}\label{myeq17}
&&[\omega-\varepsilon_{1/2\uparrow}(k_x)]\ll c_{k_x1/2\uparrow},
c_{k_x1/2\uparrow}^+(t_1) \gg_{\omega}^r\nonumber\\
&&=1\mp \varepsilon_R \ll c_{k_x2/1\downarrow},c_{k_x1/2\uparrow}^+(t_1)\gg_{\omega}^r\nonumber\\
&&+\frac{1}{2}e^{i\Omega t_1}[\ll c_{k_x2/1\downarrow},
c_{k_x1/2\uparrow}^+(t_1)\gg_{\omega+\Omega}^r\nonumber\\
&&+\ll c_{k_x2/1\uparrow},c_{k_x1/2\uparrow}^+(t_1)\gg_{\omega+\Omega}^r]\nonumber\\
&&+\frac{1}{2}e^{-i\Omega t_1}[\ll c_{k_x2/1\downarrow},
c_{k_x1/2\uparrow}^+(t_1)\gg_{\omega-\Omega}^r\nonumber\\
&&+\ll c_{k_x2/1\uparrow},c_{k_x1/2\uparrow}^+(t_1)\gg_{\omega-\Omega}^r],
\end{eqnarray}
\begin{eqnarray}\label{myeq18}
&&[\omega-\varepsilon_{1/2\downarrow}(k_x)]\ll c_{k_x1/2\downarrow},
c_{k_x1/2\downarrow}^+(t_1)\gg_{\omega}^r\nonumber\\
&&=1\pm\varepsilon_R\ll c_{k_x2/1\uparrow},c_{k_x1/2\downarrow}^+(t_1)\gg_{\omega}^r\nonumber\\
&&+\frac{1}{2}e^{i\Omega t_1}[\ll c_{k_x2/1\uparrow},
c_{k_x1/2\downarrow}^+(t_1)\gg_{\omega+\Omega}^r\nonumber\\
&&+\ll c_{k_x2/1\downarrow},c_{k_x1/2\downarrow}^+(t_1)\gg_{\omega+\Omega}^r]\nonumber\\
&&+\frac{1}{2}e^{-i\Omega t_1}[\ll c_{k_x2/1\uparrow},
c_{k_x1/2\downarrow}^+(t_1)\gg_{\omega-\Omega}^r\nonumber\\
&&+\ll c_{k_x2/1\downarrow},c_{k_x1/2\downarrow}^+(t_1)\gg_{\omega-\Omega}^r],
\end{eqnarray}
for spin-up and spin-down, respectively. It is seen from Eqs.(17)
and (18) that the retard NGF $G^r_0$ with frequency $\omega$ are
coupled to the components with photon sidebands frequencies of
$\omega+\Omega$ and $\omega-\Omega$ in connection with $k_{so}$
(the characteristic wavevector of Rashba SOC).

On the other hand, the self-energy describing the influence of the
leads on the system can be simply written as
\begin{equation}\label{myeq19}
\Sigma_{nn'}\equiv\Sigma^{L/R}_{nn'}(\omega)=2\pi\sum_{k,k_x,s}
(T^{\alpha}_{kk_xns})^*T_{k,k_xn's}^{\alpha}\delta(\omega-\varepsilon_{ks\alpha}),
\end{equation}
with which one can construct the GF $G^r=[(G^r_0)^{-1}-i\Sigma]^{-1}$ for
the whole system. If we calculate the time-averaged NGF up to the
second order, then at low temperatures the time-averaged DOS is
\begin{equation}\label{myeq20}
DOS=-\frac1{\pi}Im[Tr(G^r(\omega,\omega))],
\end{equation}
and the conductance has the form of Landauer-type$^{22,26}$
\begin{equation}\label{myeq21}
G=\frac{e^2}{h}Tr[\Sigma^L(\omega)G^a(\omega,\omega)
\Sigma^R(\omega) G^r(\omega,\omega)].
\end{equation}
Here $G^r(\omega,\omega)$ and $G^a(\omega,\omega)$ represent the
time-averaged retarded and advanced GFs, respectively.

\section{Numerical Results and Discussions}
In the following, we present some numerical examples of the DOS
and conductance calculated according to Eqs. (15)-(21) for the
system. We have selected that the energy unit
$E^*=\epsilon_1=\pi^2\hbar^2/(2m^*a^2)$ (i.e., the first lateral
level of the quantum wire without SOC), the time unit
$t^*=\hbar/E^*$, and the frequency unit $\Omega^*=1/t^*$. With
these units, the propagating longitudinal wavevector corresponding
to the $n$th transverse mode is $k_x=(\omega-n^2)^{1/2}$. In the
wide-band approximation the real part of the self-energy is
negligible,$^{22,24-26}$ and we simply assume that
$\Sigma_{11}=\Sigma_{22}=0.1$ and $\Sigma_{12}=\Sigma_{21}=0.05$.
The choice of these typical parameters is based on the following
consideration.$^{22}$ Usually the strength of electron-photon
interaction depends on the photon intensity, polarization and the
size of the quantum wire. Under the irradiation of a strong laser
with an electric field of the order $(10^5-10^6)$ $V/m$, the MEs
are comparable to or several times larger than the level spacing
in the quantum wire with the width of order $(10-100)$ $nm$
(corresponding to the external laser frequency $\sim$ THz), and
these quantities are physically realizable in recent
experiments.$^{18,19}$

We first consider the electronic structure of the system. It is
common known that the electronic energy spectrum is degenerate for
the two spin orientation in the absence of SOC. In the presence of
SOC the energy spectrum (7) satisfies the condition
$\varepsilon_{n,s}(k_x)=\varepsilon_{n,-s}(-k_x)$ in accordance
with the time inversion symmetry. However, our interest is the
interplay effect of the external laser field and the Rashba SOC on
the electronic structure and transport of the system. Here we
consider that the incident field is linearly polarized
perpendicular to the current direction (the wire direction), i.e.,
the off diagonal MEs dominate the electron-photon interaction.
With the assumption of the off diagonal MEs $V_{12}=V_{21}$=1.0
[see Eq.(11)] and the incident laser frequency $\Omega=0.5$, in
Fig.1 we illustrate the time-averaged DOS as a function of energy
for the two different Rashba SOC strengths $k_R=1/(2\pi)$ and
$k_R=1/\pi$, respectively. We can see that the main peak around
$\omega\sim1$ is always obvious in the presence of both Rashba SOC
and laser field. This is because that the electrons are populated
at energy level $\varepsilon_{1\uparrow}\sim$ 1.01 rather than
$\varepsilon_{1\downarrow}\sim$ 1.25 with single photon
absorption. In the case of weak Rashba SOC strength as shown in
Fig.1(a), there are two additional photon resonance peaks at
$\omega$=1.6 and 4.6 for spin-up (solid line), while for spin-down
(dashed line) there are three additional resonance peaks at
$\omega$=4.58, 0.75 and 0.65 with a pattern of oscillation in the
range of $0.76<\omega<1$.
\begin{figure}
\center
\includegraphics[width=1.5in]{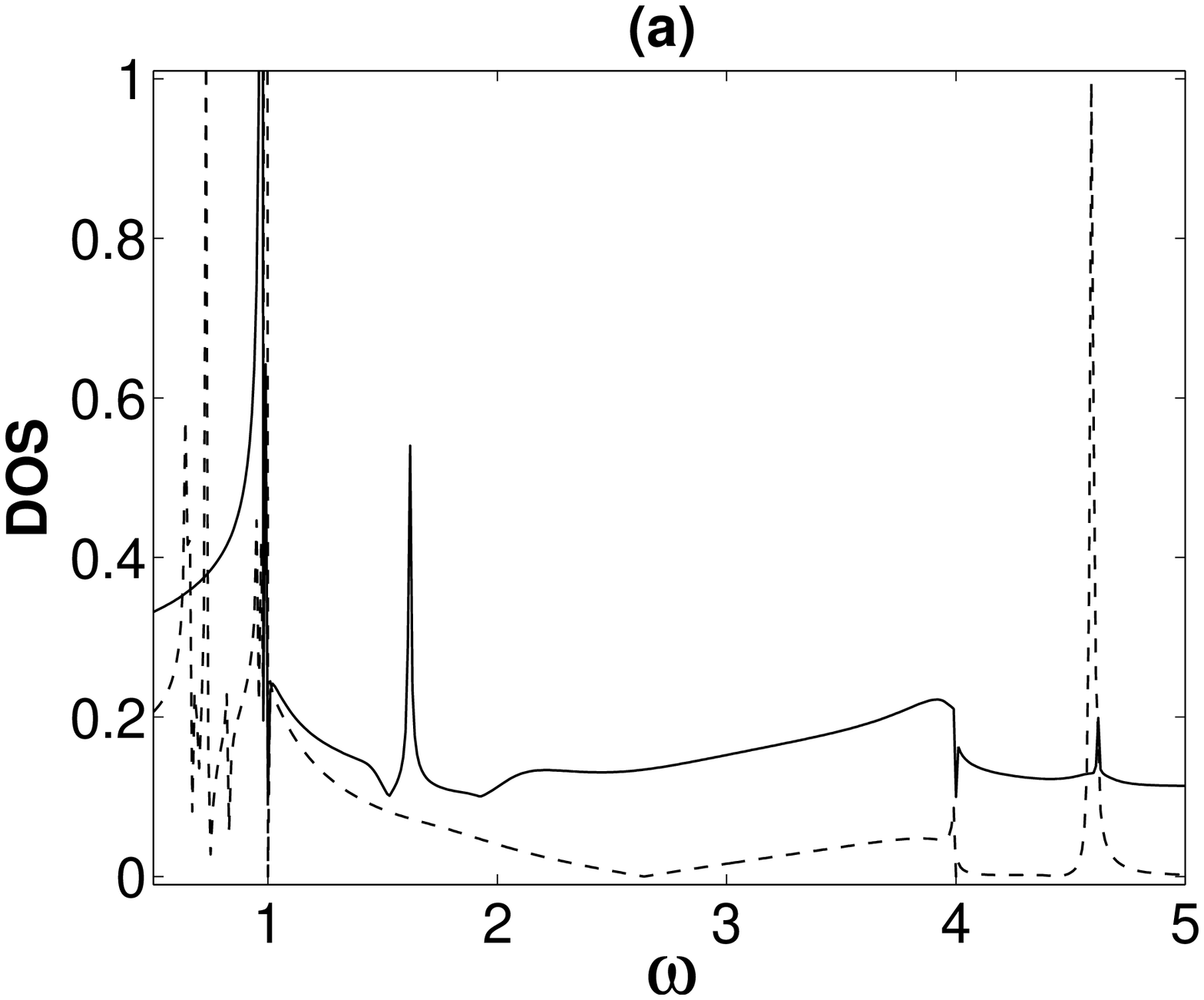}
\includegraphics[width=1.5in]{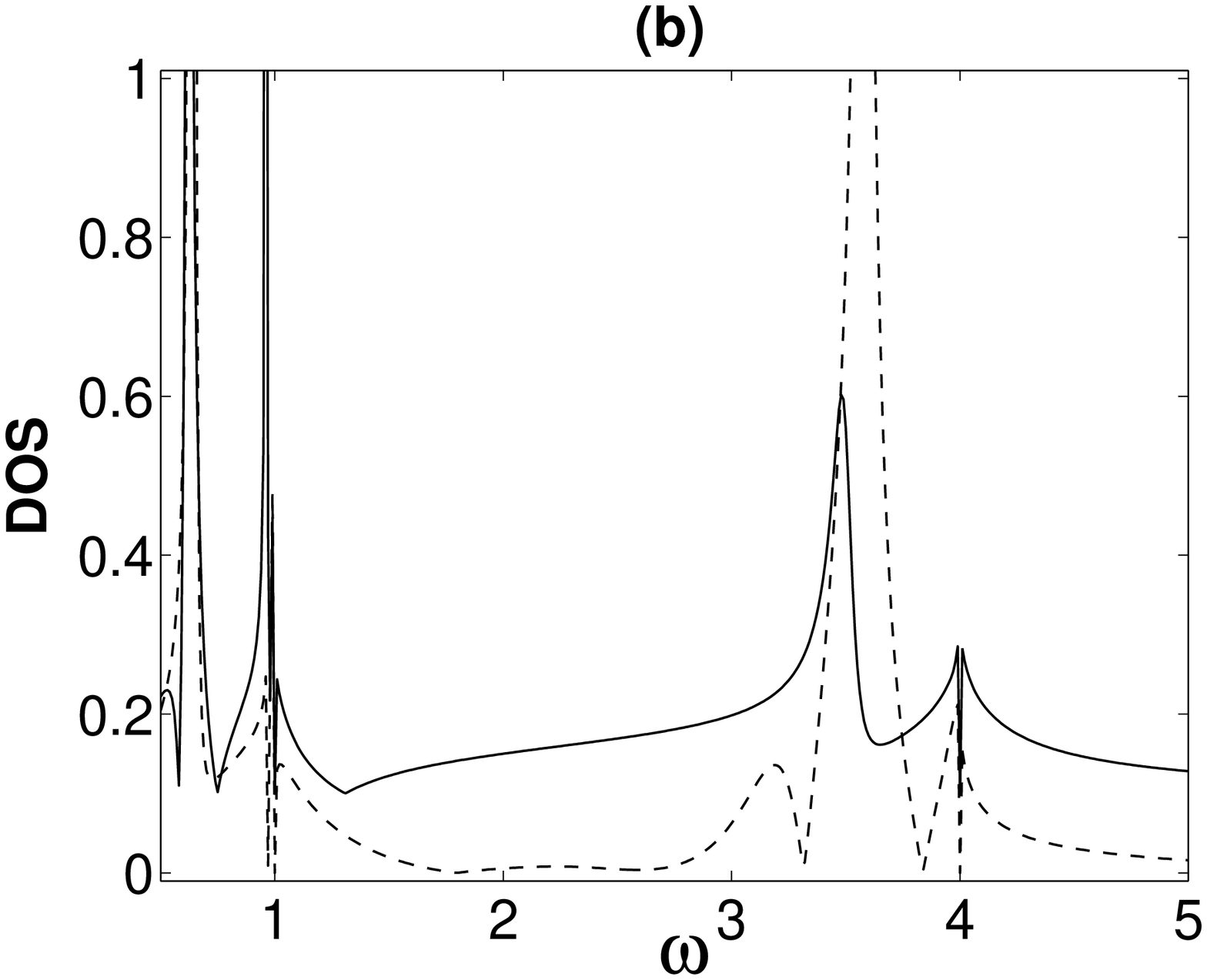}
\renewcommand{\figurename}{FIG.}
\caption{The time-averaged DOS (in arbitrary units) as a function
of energy with electron-photon interaction off-diagonal matrix
elements $V_{1s2s'}=V_{2s1s'}=1.0$ for the two different Rashba
SOC strengths (a) $k_R=1/(2\pi)$ and (b) $k_R =1/\pi$, where the
incident laser frequency is $\Omega=0.5$ and the solid (dashed)
line represents the spin-up (-down) is shifted 0.1 upward for
clarity.}
\end{figure}
\begin{figure}
\center
\includegraphics[width=1.5in]{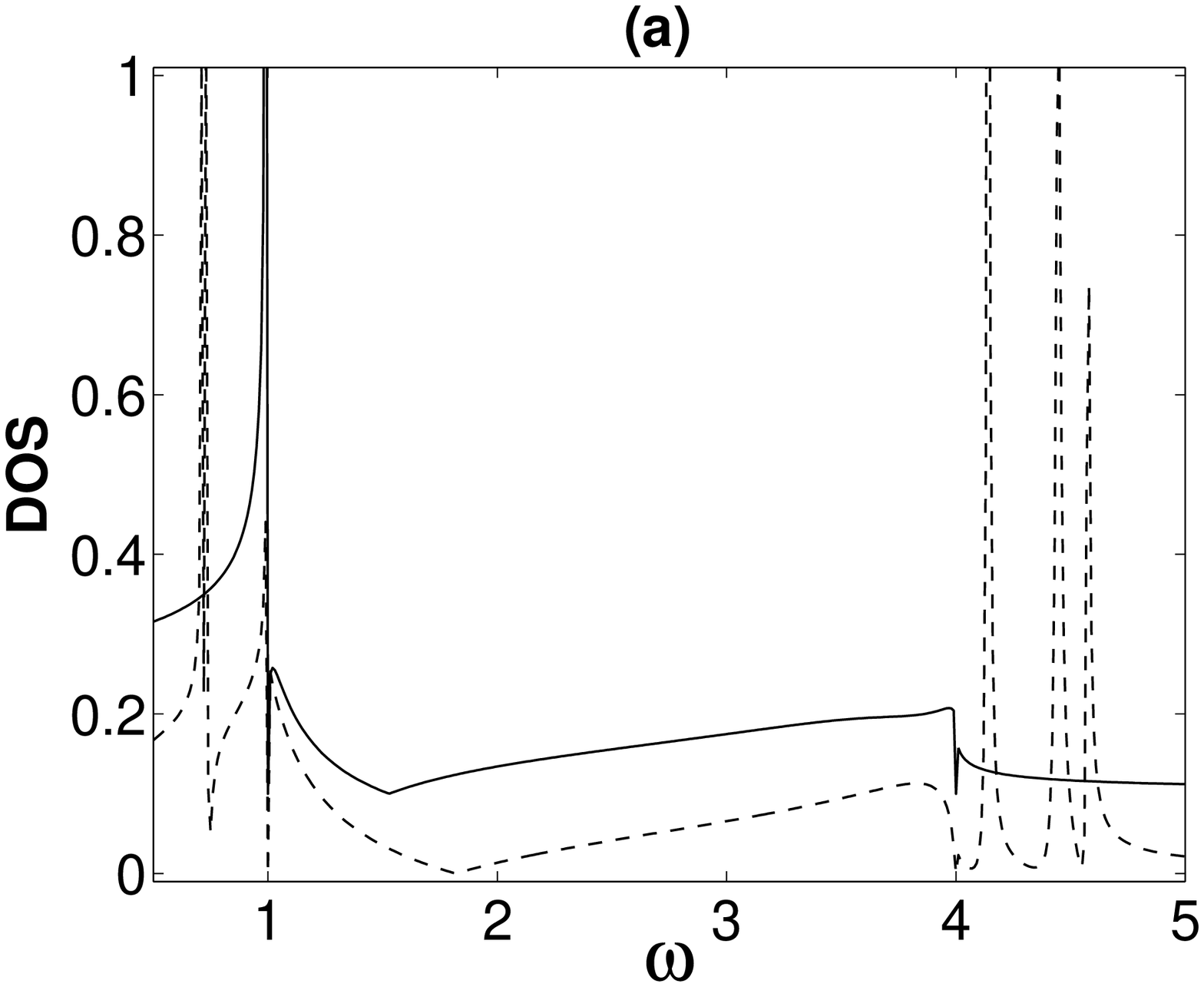}
\includegraphics[width=1.5in]{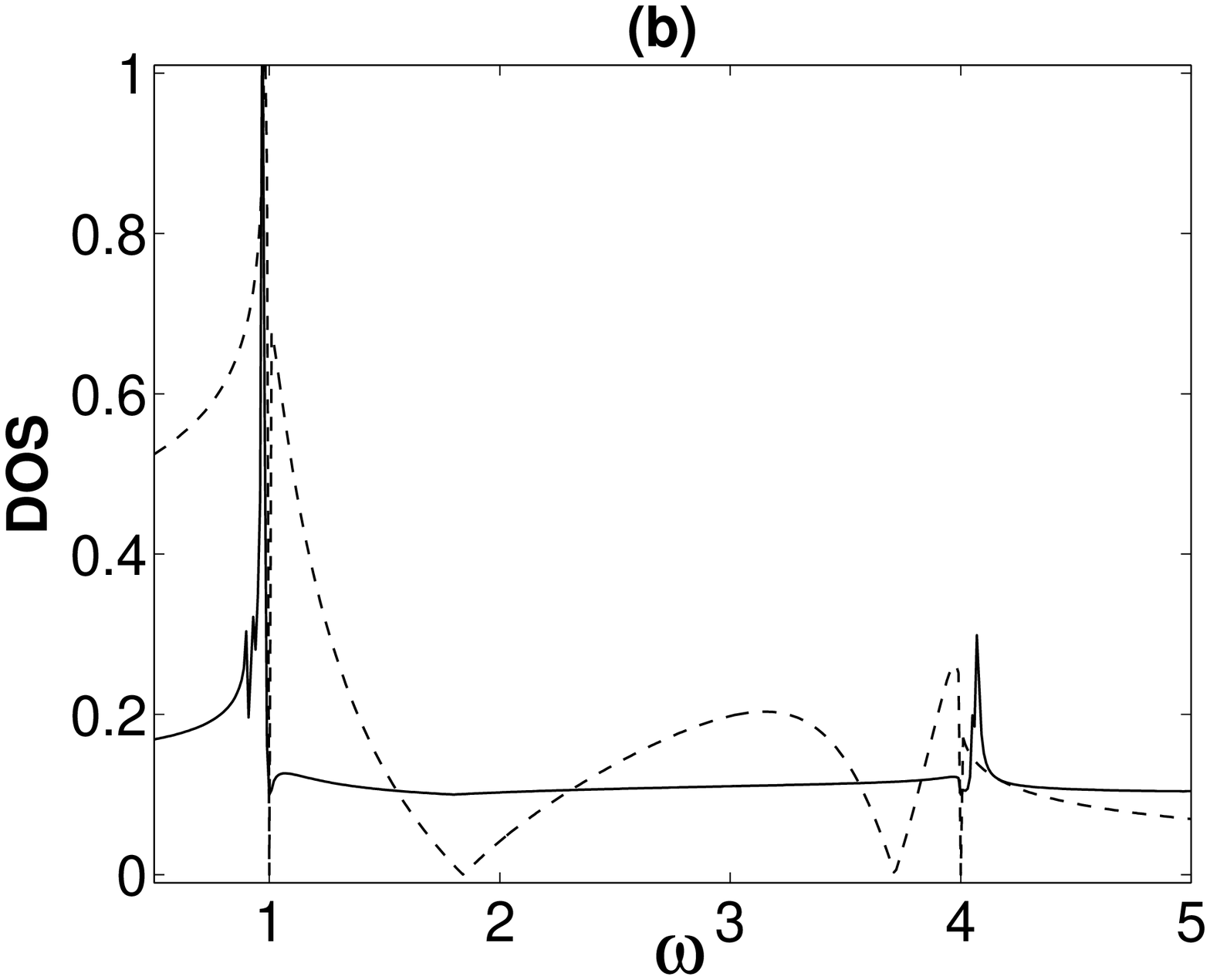}
\renewcommand{\figurename}{FIG.}
\caption{The time-averaged DOS (in arbitrary units) as a function
of energy with the same system parameters and line presentation as
in Fig.2 except for the incident laser energy is $\Omega=3.0$.}
\end{figure}
Nevertheless, as the increase of the
Rashba SOC strength shown in Fig.1(b), for spin-up the two photon
resonance peaks are shifted from $\omega$=1 and 4 to $\omega$=0.63
and 3.4, respectively. While for spin-down there only two
resonance peaks occur at $\omega=0.63$ (superposed with that for
spin-up) and 3.5 without an oscillatory pattern. However, it seems
that the other main peak around $\omega\sim4$ makes sense in this
strong Rashba SOC case. Because the single photon energy $\Omega$
is much smaller than the quantum wire sublevel spacing
$\Delta\epsilon$, the resonance peaks here are belong to the
transitions between Rashba SOC-splitted subbands with the same
lateral confinement quantum number.$^{21}$

In order to determine the transitions between subbands with
different confinement quantum number, in Fig.2 we increase the
incident frequency to $\Omega=3$ but with the same two different
Rashba SOC strengths as in Fig.1. As shown in Fig.2 the
time-averaged DOS for spin-up (solid lines) has no transition
resonance peaks in the both weak and strong Rashba SOC cases,
while for spin-down there are several sharp resonance transition
peaks at $\omega$=0.75, 4.1, 4.5 and 4.6 in the weak Rashba SOC
case [see the dashed line in Fig.2(a)] and an oscillatory pattern
with no resonance peak [dashed line in Fig.2(b)] in the strong
Rashba SOC case. This result implies a rule of possible transition
that the transition probabilities are very larger for this
condition. We believe that some of the resonance peaks in Fig.2(a)
can be identified to the photon-induced transitions between
subbands with different quantum numbers.$^{21-23}$ Because both
spin-flip and spin-conserving transitions are modulated by the
strengths of Rashba SOC and laser field, so it seems that the
strong strength of Rashba SOC in the higher laser frequency case
is not favorable for the transitions between subbands with
different quantum numbers.

\begin{figure}
\center
\includegraphics[width=1.65in]{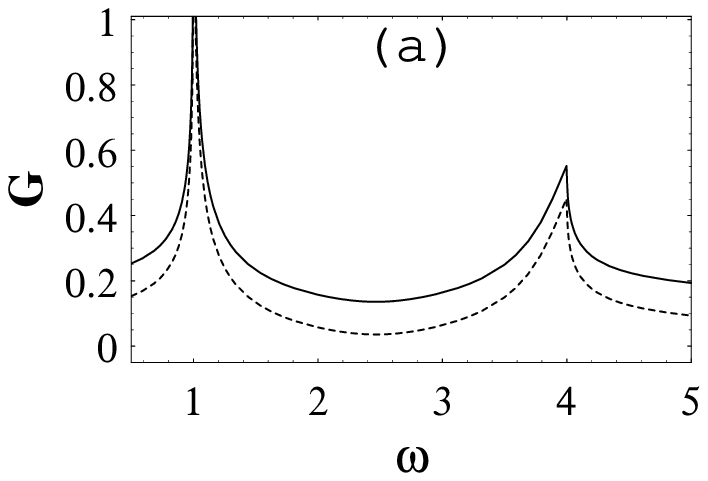}
\includegraphics[width=1.65in]{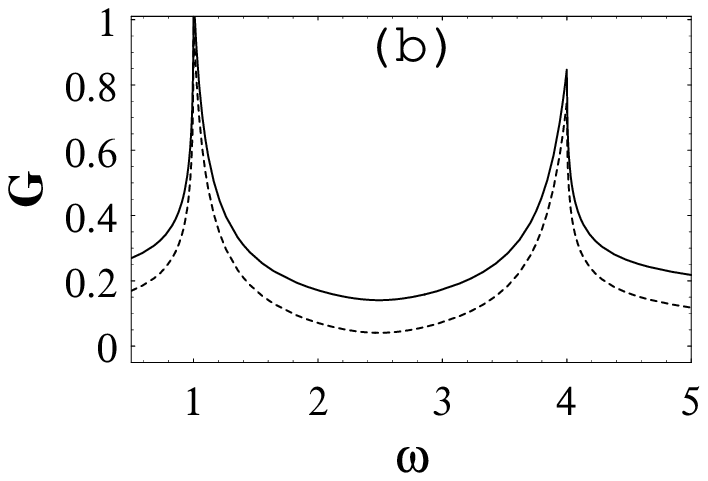}
\renewcommand{\figurename}{FIG.}
\caption{The plotted conductance $G$ (in the unit of $e^2/h$) as a
function of energy ($\sim\omega$, in unit of $\epsilon_1$) without
laser field for the two different Rashba SOC strengths (a)
$k_R=1/(2\pi)$ and (b) $k_R =1/\pi$, where the solid (dashed) line
represents the spin-up (-down) is shifted 0.1 upward for clarity.}
\end{figure}
Next we turn our attention to the conductance of the system. The
conductance (in unit of $e^2/h$) as a function of energy
($\sim\omega$, in unit of $\epsilon_1$) of the system without
external laser field in the presence of weak and strong Rashba SOC
is illustrated in Fig.3. There are two major peaks in the
conductance curves, as a consequence of the two subband levels
structure of the wire. Particularly, the conductance difference
for the two spin orientation in Fig.3 is very small and consistent
with the analytical prediction from energy spectrum. One also note
that the conductance peaks are asymmetry near the two subband
levels due to the spin-orbit interaction.$^{26}$

\begin{figure}
\center
\includegraphics[width=1.5in]{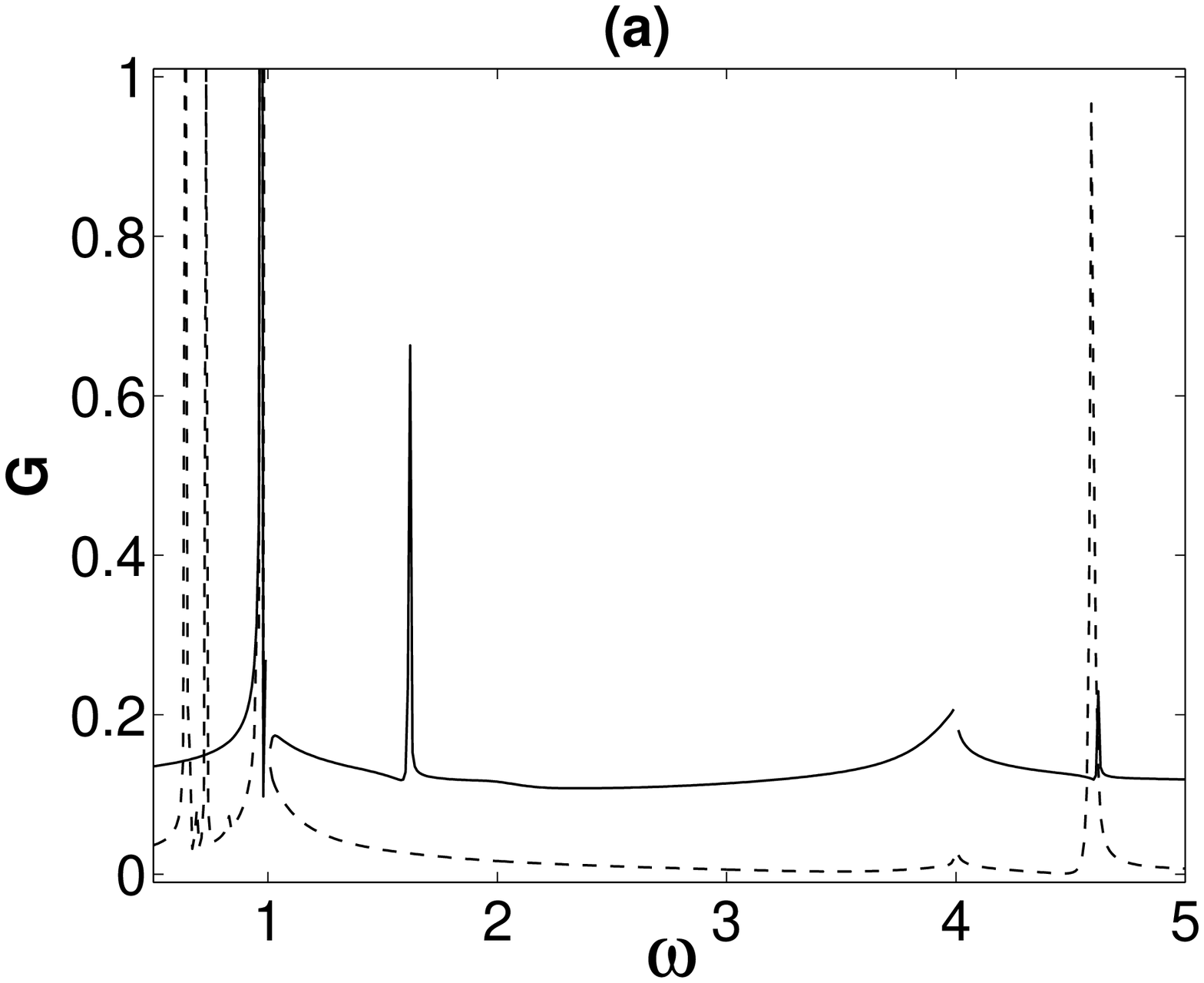}
\includegraphics[width=1.5in]{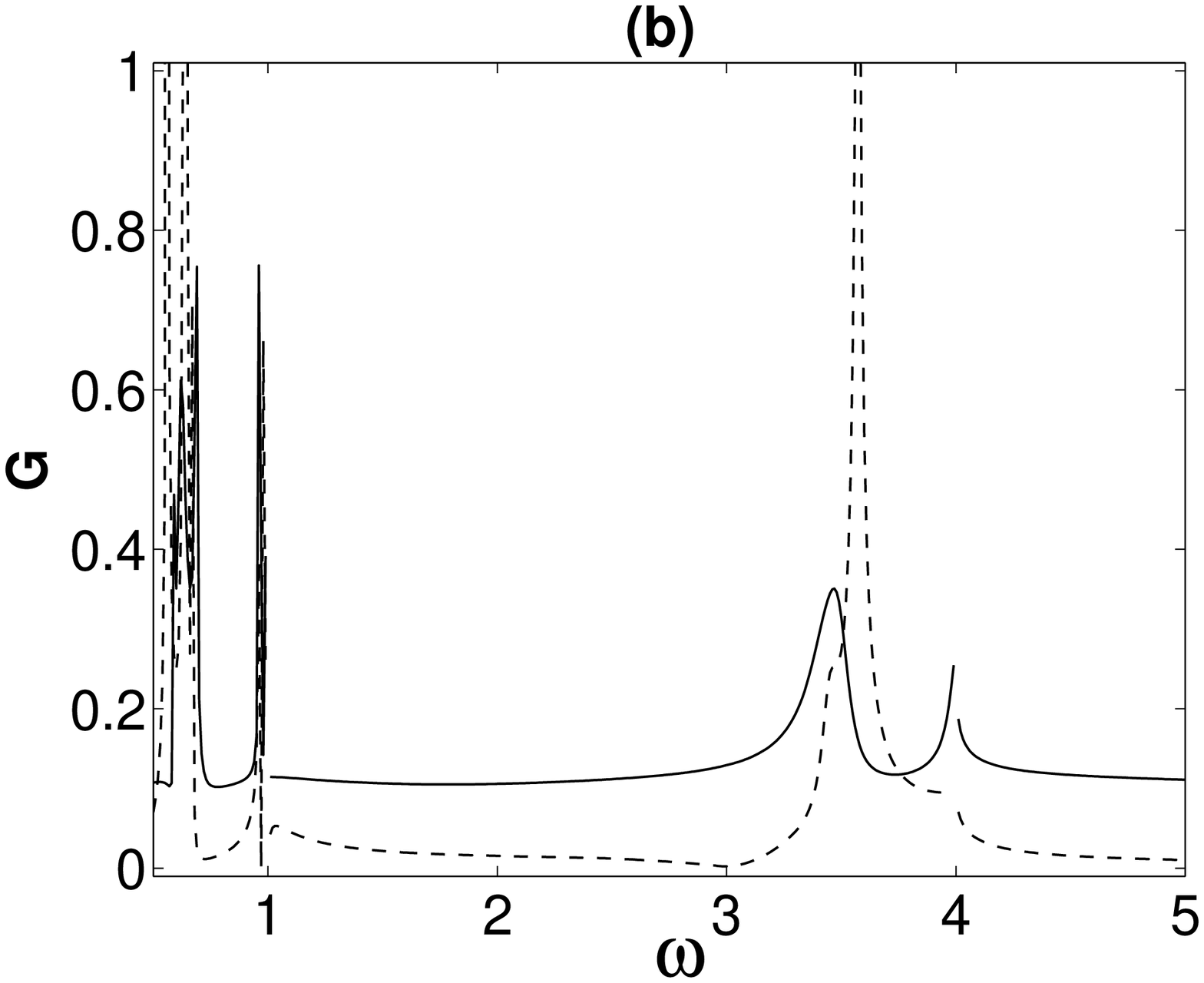}
\renewcommand{\figurename}{FIG.}
\caption{The time-averaged conductance $G$ (in the unit of
$e^2/h$) as a function of energy with the same system parameters
and line presentation as in Fig.1.}
\end{figure}
The time-averaged conductance of the system irradiated under a
transversally polarized laser field in the presence of Rashba SOC
is shown in Fig.4 with $\Omega=0.5$. Corresponding to the
resonance states in Fig.1(a), the time-averaged conductance in
Fig.4(a) shows some peaks with the height of $\sim e^2/h$. When
the incident electrons energy is about $\omega=$0.65 and 0.75, we
note that the conductance is nearly $e^2/h$ for spin-down while
that for spin-up is nearly 0; when the incident electrons energy
is increased to $\omega=1.6$, there is a sharp conductance peak
for spin-up while that for spin-down is about 0. Therefore, with a
largest spin polarization in Fig.1(a), a spin filter may be
devised in the case of appropriate incident electron energy and
the Rashba SOC strength. Fig.4(b) shows the time-averaged
conductance corresponding to the Fig.1(b) in strong Rashba SOC
case, from which one can see more photon resonance peaks
(especially in lower energy range) than in the weak Rashba SOC
case. Furthermore, when the external laser frequency is increased
to 3.0 the time-averaged conductance of the system with the two
different Rashba SOC strengths is illustrated in Fig.5. Due to the
intersubband resonance states in Fig.2(a), there is more sharp
resonance transition peaks in higher energy range [see Fig.5(a)]
for the spin-down electrons [see the explanation for Fig.2(a)].
While in the strong Rashba SOC case, the conductance curves for
both spin-up and -down show only the two main peaks [see Fig.5(b)]
as in Fig.2(b). Maybe in this case the Rashba SOC is too
strong to produce quantum transitions for the system.
\begin{figure}
\center
\includegraphics[width=1.5in]{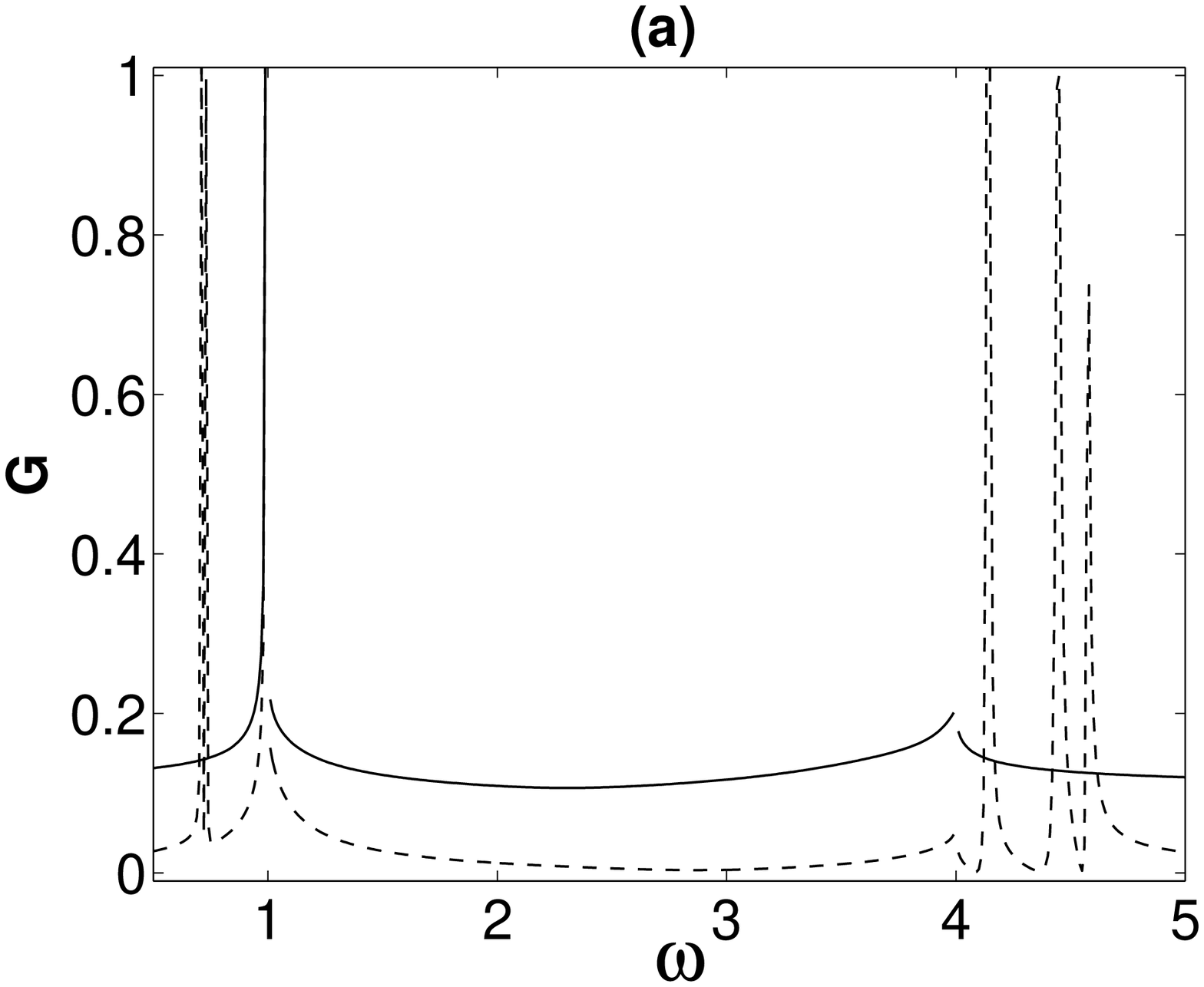}
\includegraphics[width=1.5in]{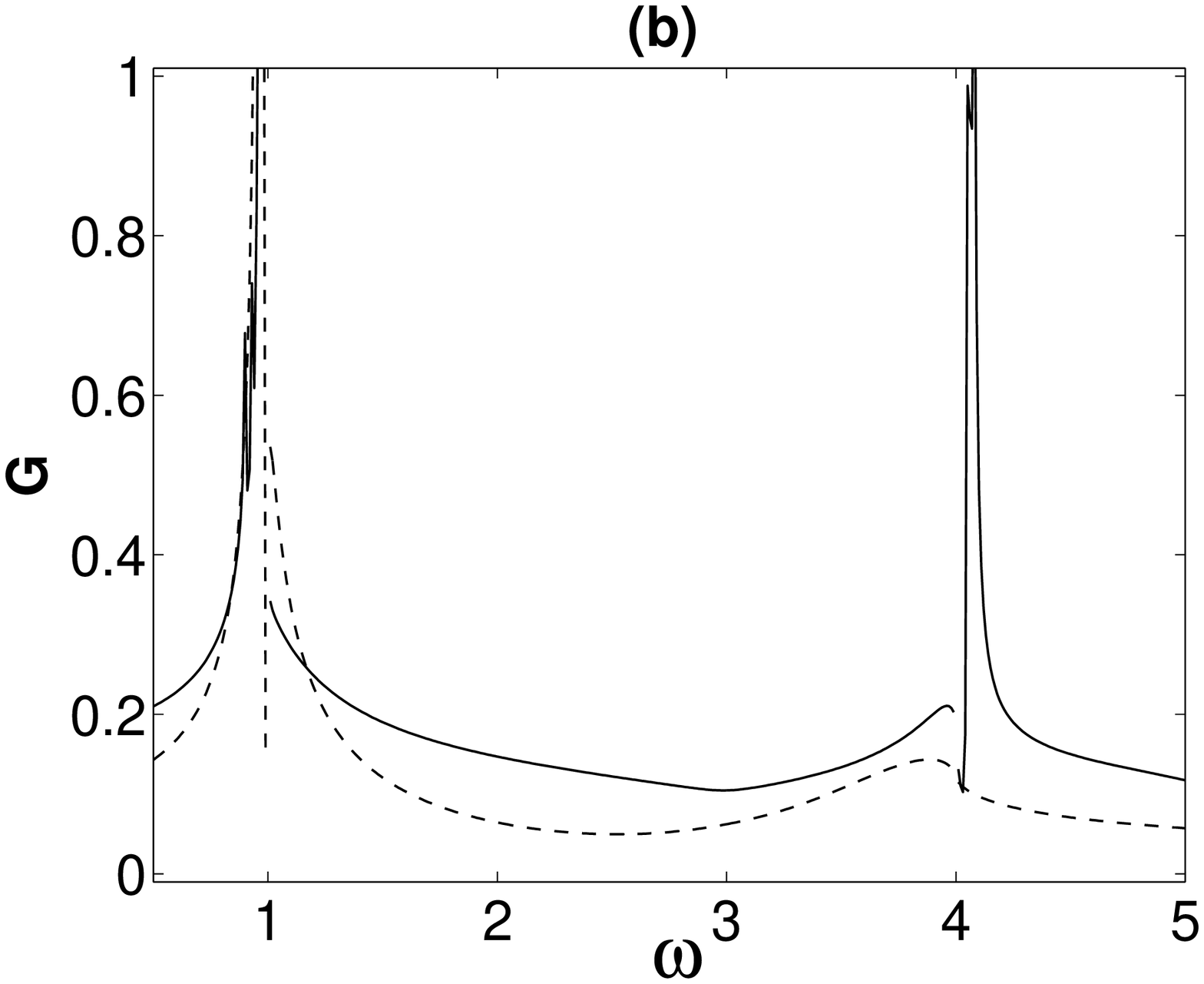}
\renewcommand{\figurename}{FIG.}
\caption{The time-averaged conductance $G$ (in the unit of
$e^2/h$) as a function of energy with the same system parameters
and line presentation as in Fig.2.}
\end{figure}

Finally, the time-averaged DOS (solid line for spin-up and
dash-dotted line for spin-down) and the spin polarization
rate$^{20}$ (dashed line) with a fixed incident electron energy
($\omega=2.5$) as a function of the characteristic wavevector
$k_R$ (proportional to the strength of Rashba SOC) without or with
a transversally polarized external laser field ($\Omega=0.5$) are
demonstrated in Fig.6(a) and Fig.6(b), respectively. The
electronic energy spectrum is degenerate for spin-up and spin-down
when $k_R=0$ in both cases as expected (see the solid and
dash-doted lines in Fig.6. In the case without laser field as
shown in Fig.6(a), the spin polarization rate (dashed line) is
about 17\% when $k_R=0.02$, and it can reach to 95\% when
$k_R=0.04$. Under the irradiation of the laser field, as shown in
Fig.6(b), the spin polarization rate increases to 60\% and 100\%
around $k_R=0.02$ and $k_R=0.04$, respectively. Moreover, there
are several additional peaks of spin polarization rate in the
range of $0.05<k_R<0.25$ with laser field, while in the case
without laser filed as shown in Fig.6(a) the spin polarization
rate is smoothly low in this range of $k_R$. Therefore, it seems
that the external laser field can enhance the spin polarization
rate for a quantum wire system with an appropriate Rashba SOC
strength which can be adjusted through the controllable lateral
electrodes.$^{16}$
\begin{figure}
\center
\includegraphics[width=1.5in]{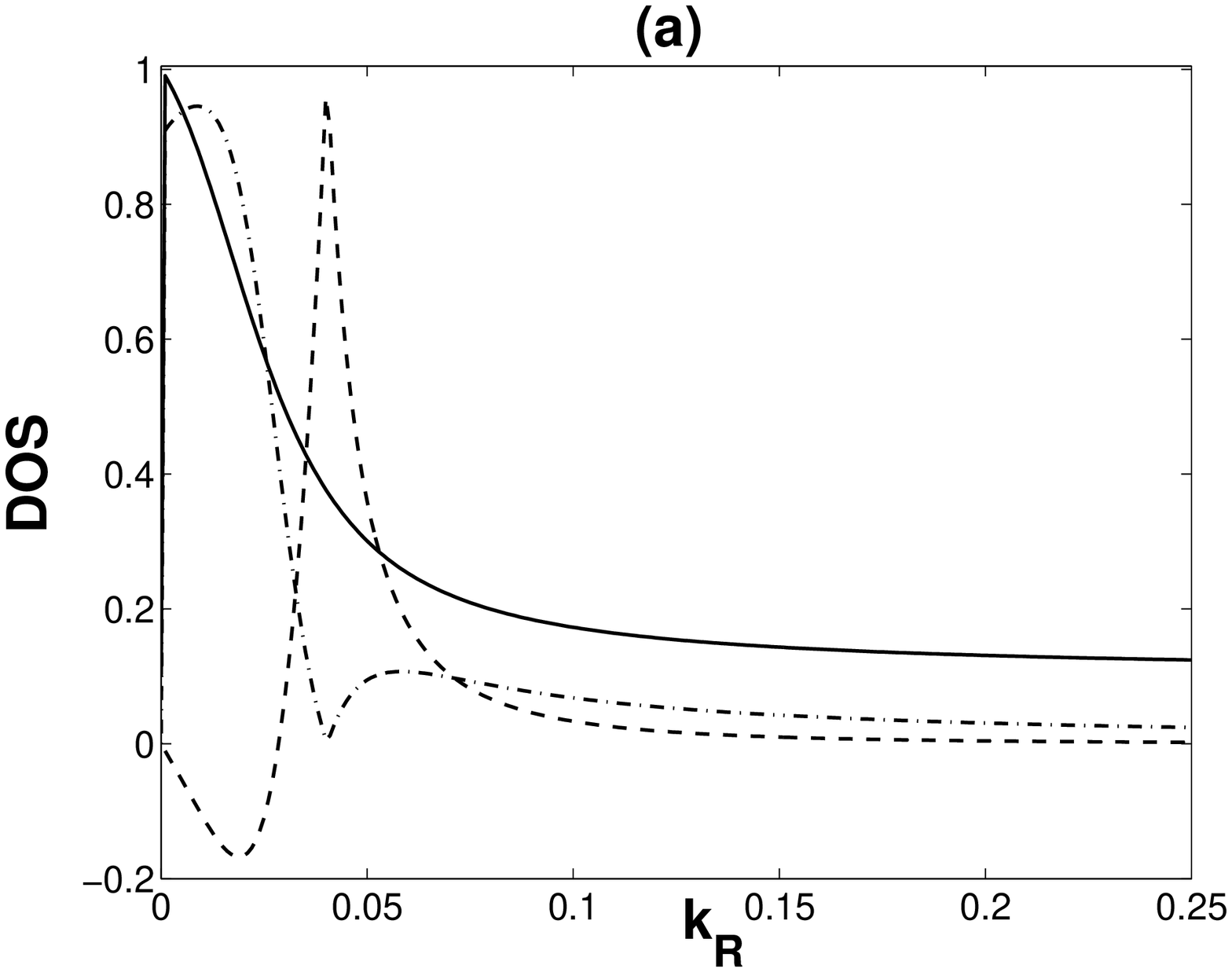}
\includegraphics[width=1.5in]{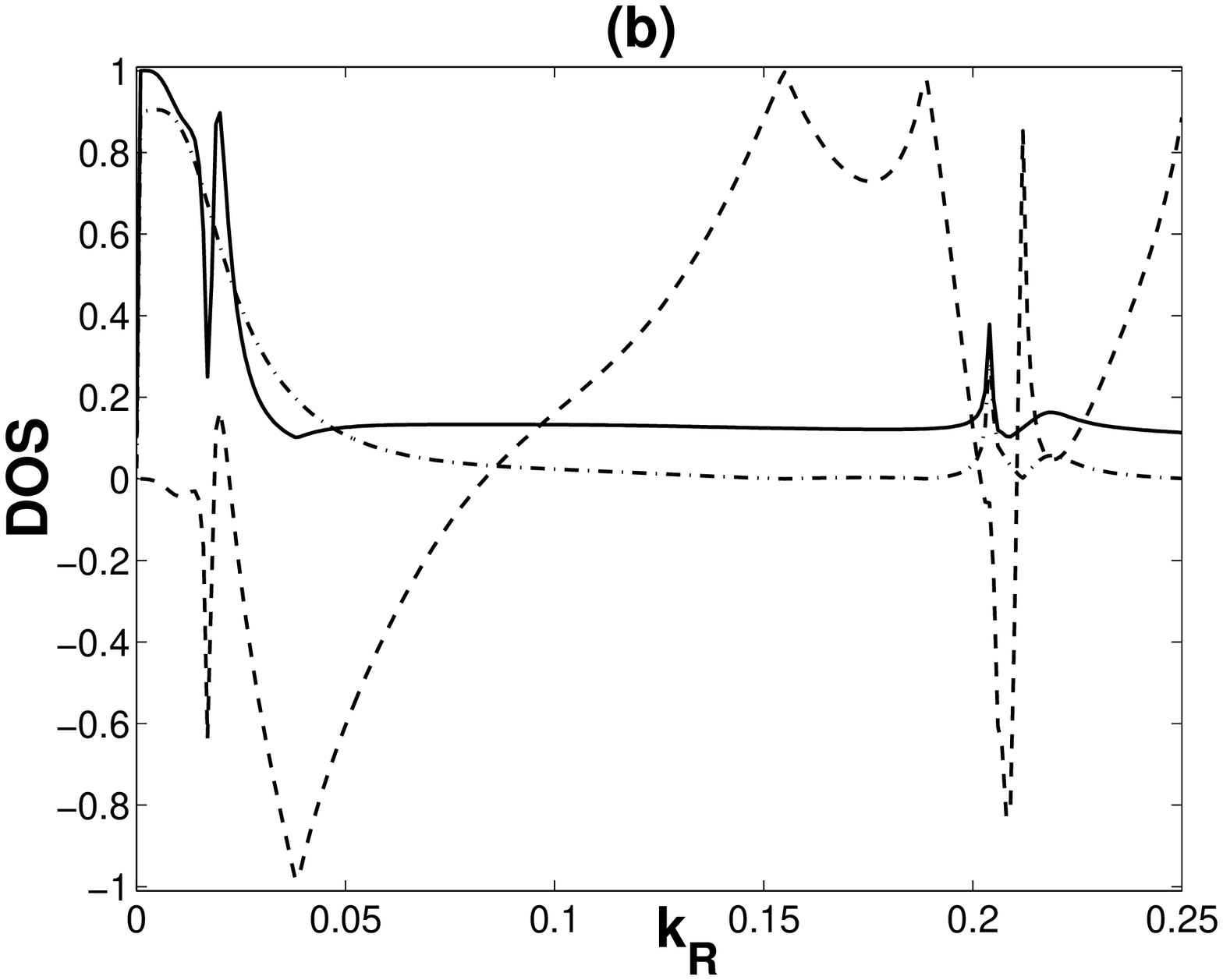}
\renewcommand{\figurename}{FIG.}
\caption{The time-averaged DOS and spin polarization rate as a
function of $k_R$ (proportional to the strength of Rashba SOC) for
a fixed incident electrons energy $\omega=2.5$ (a)without and
(b)with a transversally polarized laser field ($\Omega=0.5$),
where the solid line (shifted 0.1 upward for clarity) for spin-up
and dash-dotted line for spin-down DOS, respectively. The dashed
line represents spin polarization rate.}
\end{figure}

\section{Conclusion}
In summary, using the method of EOM for Keldysh NGF, we have
investigated theoretically the electronic structure and transport
properties of a two-sublevel quantum wire irradiated under a
transversally polarized external laser field in the presence of
Rashba SOC. The time-averaged DOS and conductance for spin-up and
spin-down electrons in the case of the off-diagonal
electron-photon interaction dominating the process are calculated
analytically, and are demonstrated numerically with two different
Rashba SOC strengths and laser frequencies, respectively. It is
found that the external laser field can enhance the spin
polarization rate for the system with some particular Rashba SOC
strengths. An all-electrical nonmagnetic spintronic devices may be
desirable under an appropriate choice of external control
parameters. However, the experimental observation for this
proposal and further theoretical investigation if the impurity,
phonon or electron-electron interaction are taken into account are
worthy to be carried out.

\begin{acknowledgements}
This work was supported by National Natural Science Foundation of
China (Grant NO. 10574042), and by Scientific Research Fund of
Hunan Provincial Education Department (Grant NO. 04A031).
\end{acknowledgements}


\end{document}